\documentclass[a4paper,11pt]{article}
\usepackage{pos}
\usepackage{gensymb}

\title{The Small-Sized Telescopes for the Southern Site of the Cherenkov Telescope Array}
\ShortTitle{SST}

\author*[a]{Richard White}
\affiliation [a]{Max-Planck-Institut f\"ur Kernphysik, P.O. Box 103980, D69029 Heidelberg, Germany}

\forColl{CTA SST} 
\emailAdd{richard.white@mpi-hd.mpg.de}

\abstract{The Cherenkov Telescope Array (CTA) will use three telescope sizes to efficiently detect cosmic gamma rays in the energy range from several tens of GeV to hundreds of TeV. The Small-Sized Telescopes (SSTs) will form the largest section of the array, covering an area of many square kilometres on the CTA southern site in Paranal, Chile. Up to 70 SSTs will be implemented by an international consortium of institutes and teams as an in-kind contribution to the CTA Observatory. The SSTs will provide unprecedented sensitivity to gamma rays above 1 TeV and the highest angular resolution of any instrument above the hard X-ray band. CTA has recently finalised the technology that will be used for the SSTs: the telescopes will be a dual-reflector design with a primary reflector of ~4 m diameter, equipped with an SiPM-based camera with full waveform readout from $\sim$2000 channels covering a $\sim$9$^\circ$  field of view. The Schwarzschild-Couder optical configuration leads to a small plate-scale, and consequently a compact, cost-efficient camera. In this contribution, we describe the experience gained operating telescope and camera prototypes during the CTA preparatory phase, and the development of the final SST design.}

\FullConference{37$^{\rm{th}}$ International Cosmic Ray Conference (ICRC 2021)\\
		July 12th -- 23rd, 2021\\
		Online -- Berlin, Germany}

\begin{document}
\maketitle

%
\section{Introduction}
\begin{figure}[b!]
    \centering
    \includegraphics[width=0.8\textwidth]{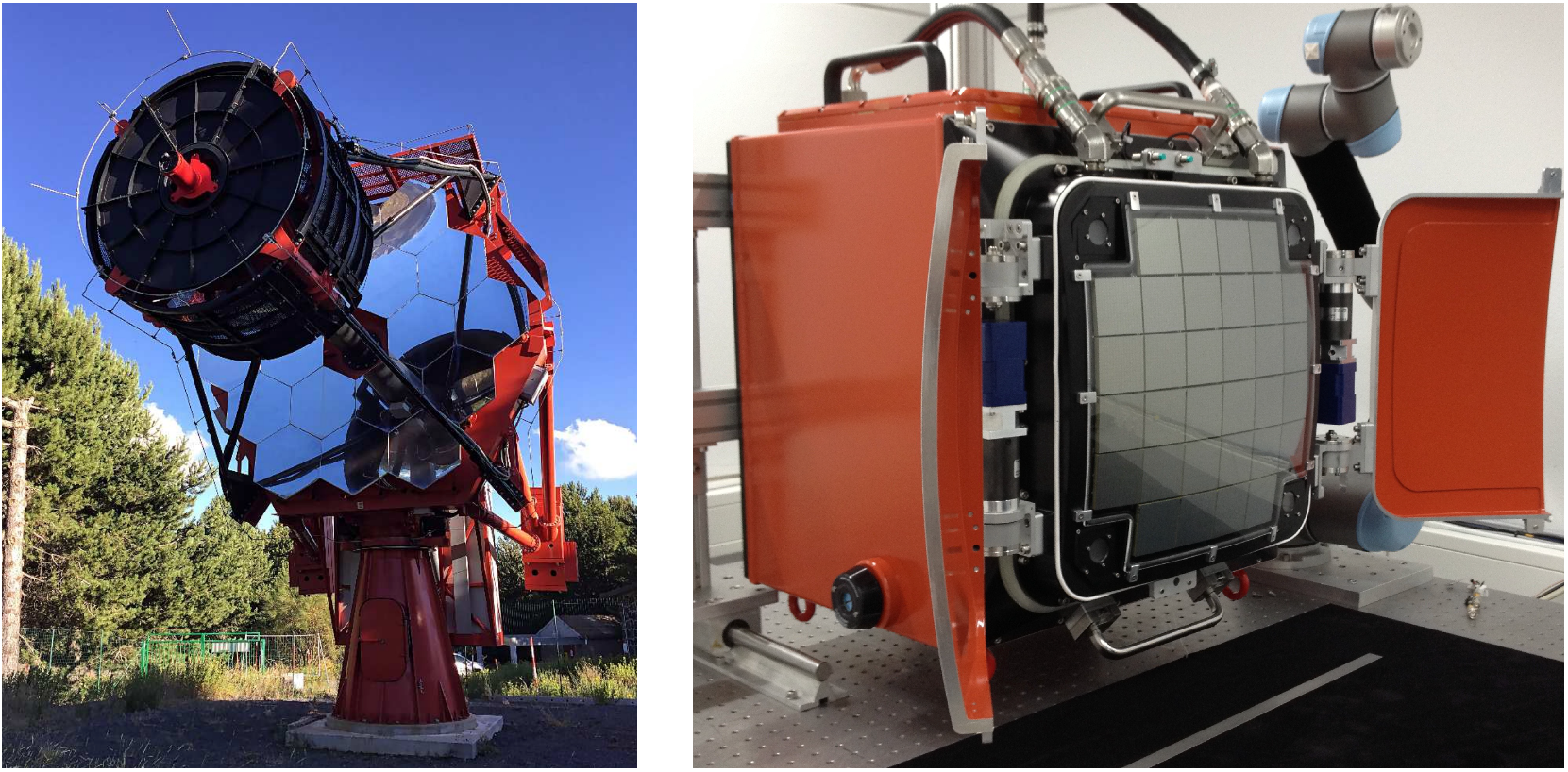}
    \caption{{\bf Left:} The ASTRI-Horn telescope prototype structure at the Catania Astrophysical Observatory (Serra La Nave, Mount Etna). {\bf Right:} The CHEC-S camera prototype undergoing laboratory tests.}
    \label{proto}
\end{figure}

The southern site of CTA (CTA-South) in Paranal, Chile will host a large number of Small-Sized Telescopes (SSTs), purposed with detecting gamma rays up to energies of several hundred TeV. The SST array offers an opportunity to provide CTA with unprecedented sensitivity and the highest angular resolution of any instrument operating above X-rays. To achieve this many telescopes are required. An optimal SST is therefore of reasonable cost, highly reliable and easily maintainable. A camera for the SST should have a large field of view (FoV) (to capture large and off-axis images), have fine pixelisation (to resolve small images and isolate signal from background) and have a large readout window (to fully contain images with a large time gradient). Prototype SST structures and cameras have been developed, built and tested by several groups (\cite{sst1m, astri-program, chec, gate}). Following a harmonisation process in 2019, a formal decision was made to base the SST on the dual-reflector design adopted by the ASTRI-Horn and CHEC-S prototypes (see Figure~\ref{proto}). A dual-reflector optical system, such as that proposed by Schwarzschild and Couder and recently re-purposed for gamma-ray astronomy~\cite{Vassiliev07}, offers good performance out to large field angles and provides the opportunity to reduce the telescope plate scale, thereby allowing the used of silicon-photomultipliers (SiPMs) as photosensors reducing costs whilst improving reliability and detection efficiency (c.f. Photomultiplier Tubes). The SST Programme has recently been established to finaslise the SST design and provide all SSTs as an in-kind contribution to the CTA Observatory. The programme will incorporate lessons learnt from all SST prototyping efforts. Critical to this are the efforts of the GATE team, who produced an alternative dual-reflector system used to detect the first Cherenkov light of any CTA prototype at the Observatoire de Paris~\cite{gct-perf, gct-perf2, checm}.

\section{Optical \& Mechanical Design and Implementation}

\begin{figure}[t!]
    \centering
    \includegraphics[width=0.97\textwidth]{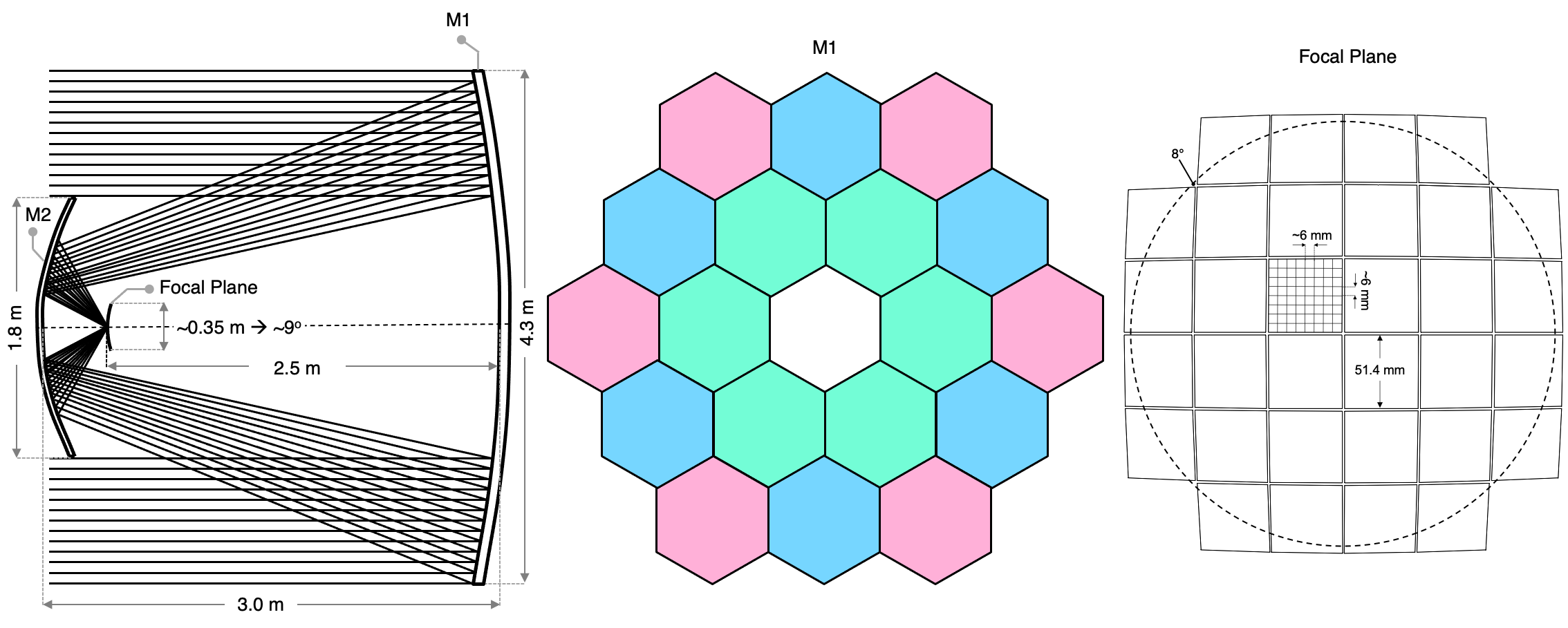}
    \caption{{\bf Left:} An overview of the SST optical system. {\bf Center:} The layout of mirrors composing the M1 reflector, where each colour indicates a mirror with a different shape. {\bf Right:} The focal plane layout indicating the FoV required by CTA (8$^\circ$) and the nominal photodetector positions.}
    \label{optics}
\end{figure}

An overview of the SST optical design is shown in Figure~\ref{optics}. The system consists of two reflectors, each following an aspherical shape described by a high-order polynomial~\cite{sironi2017}. The resulting focal plane surface is curved with a spherical radius of 1060~mm. 

The telescope structure has a primary mirror (M1) diameter of 4.3~m, a primary-to-secondary distance of 3~m and a distance from the camera to the secondary mirror (M2) of 0.52~m. The structure is an altitude-azimuth design in which the azimuth axis permits a rotation range of $\pm$270$^\circ$. The mirror dish is mounted on the azimuth fork, which allows rotation around the elevation axis from $-$1$^\circ$ to $+$90$^\circ$. The mast structure that supports the secondary mirror and the camera is fixed on the mirror dish. The resulting structure has a weight of 16~tons, after optimisation, providing a high-level of mechanical stiffness advantageous for mirror alignment, telescope pointing, and minimising risks presented by extreme environmental conditions such as earthquakes in Chile, being able to move to safe position after tremors. 

Eighteen hexagonal, low area density ($\sim$10~kg/cm$^2$) panels make up the primary surface~\cite{Canestrari2014}. To reproduce the aspherical optical profile of M1, the surface is arranged in three concentric rings each built from mirror segments with different profiles. The M2 reflector is a monolithic substrate thick glass shell of 19~mm thickness and 1.8~m diameter bent to the desired curvature. The resulting optical system has a plate scale of 37.5~mm/$^\circ$, an angular pixel size of 0.19$^\circ$, and an equivalent focal length of 2150~mm. This setup delivers a FoV up to 10.5$^\circ$ in diameter and a mean value of the effective area of about 6~m$^2$, ensuring that more than 80\% of the light emitted by a point source is collected within the dimensions of a Cherenkov camera pixel over the full FoV of the telescope.

\begin{figure}[t!]
    \centering
    \includegraphics[width=0.95\textwidth]{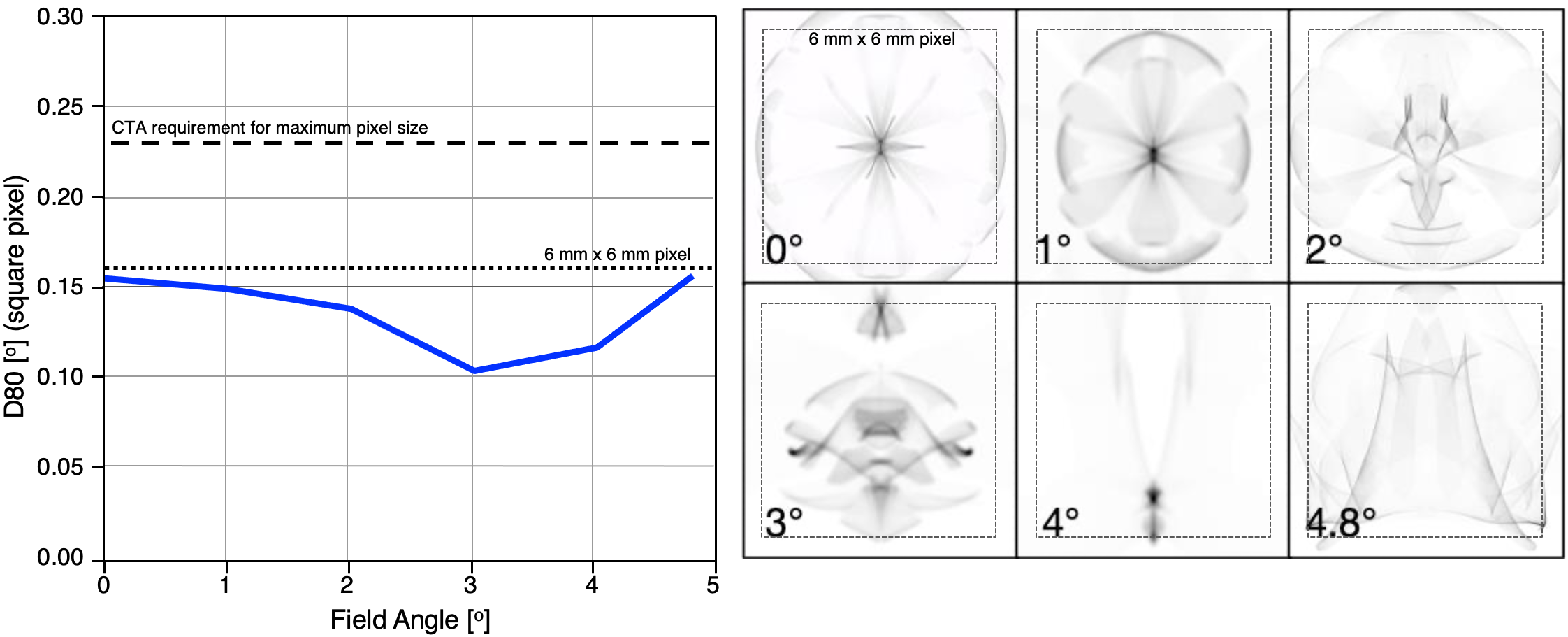}
    \caption{{\bf Left:} Square containing 80\% of enclosed energy (D80) vs. off-axis angle (solid line), CTA requirement for the maximum pixel size (dashed line, 0.23$^\circ$) and SST pixel size (dotted line, 0.16$^\circ$). {\bf Right:} Simulated images resulting from point source located 10~km above the telescope for various field angles.}
    \label{psf}
\end{figure}

The ASTRI-Horn prototype telescope structure (Figure~\ref{proto}, left) was developed by the Italian National Institute of Astrophysics, (INAF). It was installed at the Catania Astrophysical Observatory (located at Serra La Nave on Mount Etna) during Fall 2014. In 2017 the prototype successfully provided a robust optical validation of a Schwarzschild-Couder telescope for the first time~\cite{astri-optical-v}. Figure~\ref{psf} shows the measured PSF as a function of off-axis angle. The performance of telescope to Cherenkov light was later demonstrated using the ASTRI camera, reporting detection of the Crab Nebula in gamma rays for the first time with a such dual-reflector imaging atmospheric Cherenkov telescope~\cite{astri-crab}. The final telescope mechanical design is under optimisation for production and deployment, including  a reduction in mass (by$\sim$ 25\%, whilst maintaining stiffness). 

\section{Camera Design and Implementation}

The SST Camera will be based on the CHEC-S prototype (see Figure~\ref{proto}, right)~\cite{checs}. The camera contains 2048 pixels tiled in the focal plane to approximate the radius of curvature resulting from the telescope optics. The underlying architecture consists of 32 identical camera modules connected to a single Backplane (BP). Data is output on a 10 Gbps interface. Each camera module contains an SiPM tile with 64-pixels each of active size 6~mm~$\times$~6~mm. Each tile is connected to Front-End Electronics (FEE) module, based on TARGET Application Specific Integrate Circuits (ASICs)~\cite{target}, that provides SiPM bias control, waveform digitisation and a first-level camera trigger. Four 16-channel CTC (CTA-TARGET-C version) ASICs provide sampling whilst four 16-channel CT5TEA (CTA-TARGET-5TEA version) ASICs are used for triggering. The CTC ASIC is a 12-bit device that provides an effective dynamic range of 1 to $\sim$500 photoelectrons as configured for the SST Camera. The recovery of larger signals off line is possible due to the waveform digitisation. The sampling rate of CTC is tuneable, but nominally set to 1~GSa/s. The size of the readout window digitised from the storage array is settable in 32~ns blocks, nominally set to 96 - 128 ns to capture high-energy, off axis events as they transit the focal plane. A slow-signal digitisation chain providing a per-pixel measurement of the DC light level is included to track the pointing of the telescope via stars during normal operation. A Field Programmable Gate Array (FPGA) on-board each FEE module is used to configure the ASICs and other components, to read-out raw data from the ASICs, and to package and buffer raw data for output from the module. The 32 camera modules are connected to the BP that provides the interface for power, clock, trigger and data. The BP forms a nanosecond-accurate camera trigger by combing signals from all FEE modules in a single FPGA (nominally set to require a coincidence between two neighbouring FEE trigger patches). Following a camera trigger a serial message containing a unique event identifier is sent to the FEE to retrieve data from the sampling ASICs at the appropriate position in their memory. Data and communication links to the FEE modules are routed via the BP off-camera via 10 Gbps fibre-optic link. The full 512-bit camera trigger pattern is also sent on an even-by-event basis. An array-wide White Rabbit system connected to a timing board inside the camera provides absolute timing. The camera includes an illumination system to provide calibration via fast, variable intensity, LED flashes. An entrance window and external door system provide protection from the elements. Thermal control of the camera is via an external chiller. Chilled liquid is circulated through the camera focal plane plate and a thermal exchange unit on the camera body. Fans internal to the camera circulate the resulting cooled air. The camera is hermetically sealed and a breather-desiccator is used to maintain an acceptable level of humidity and atmospheric pressure. A overview of the current SST Camera CAD is shown in Figure~\ref{cam}. The design is in-progress, with focus on verifying several critical design changes from CHEC-S:

\begin{figure}[t!]
    \centering
    \includegraphics[width=0.95\textwidth]{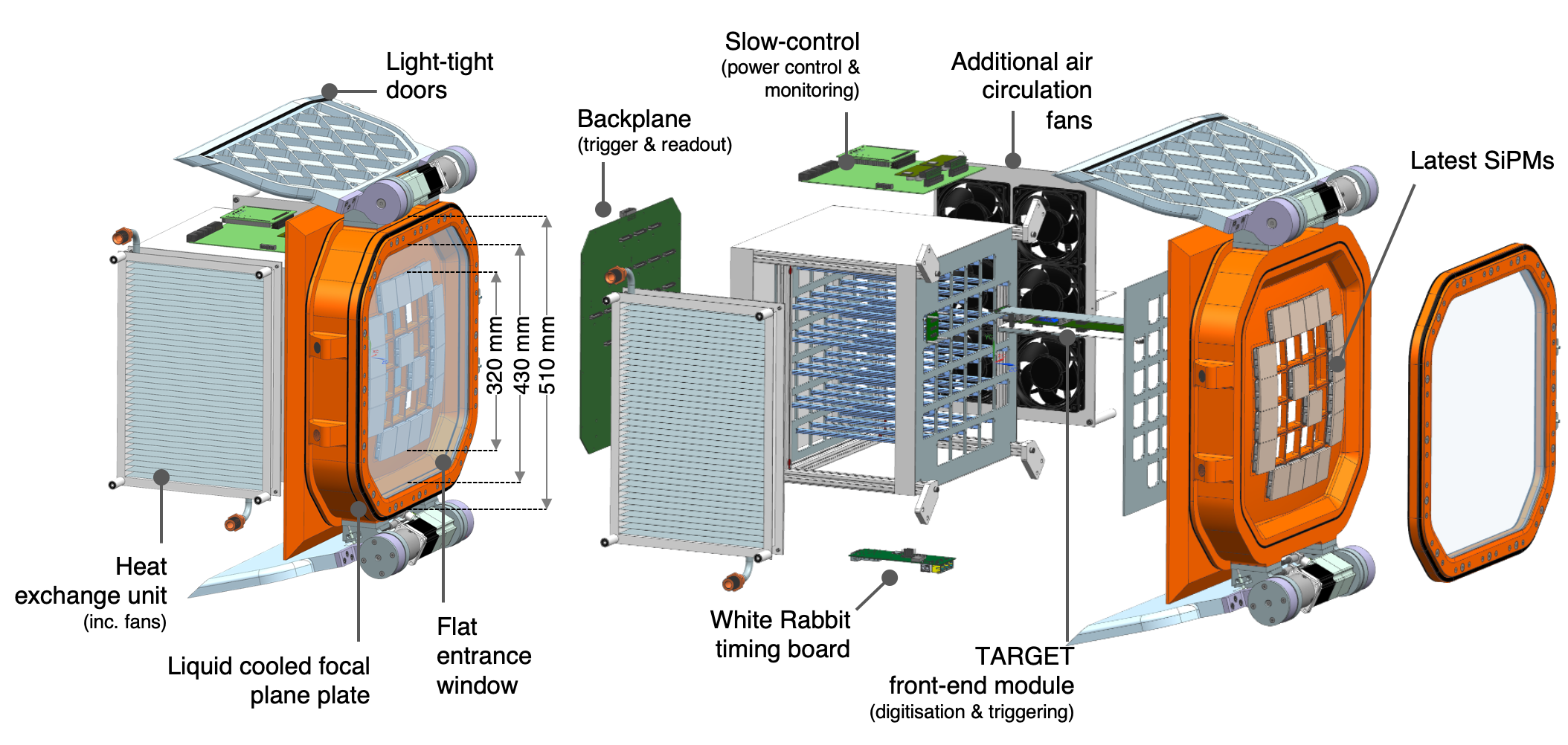}
    \caption{An overview of the camera CAD indicating the major sub-systems.}
    \label{cam}
\end{figure}

\begin{itemize}
  \item {\bf Entrance window:} The curved CHEC-S window is replaced with a flat version for ease of manufacture. To prevent obscuring the outer pixels on the curved focal plane, the window must extend past the SiPM array (increasing the cameras size, though the effect on telescope shadowing / effective collection area is negligible). The entrance window is coated to optimise the transmission of Cherenkov light whilst attenuating the night-sky background (NSB). 
  \item {\bf SiPMs:} The latest SiPMs offer improved photon detection efficiency and lower optical cross talk than those used in CHEC-S. In addition work is taking place with the proposed manufacture to mount pixels onto a carrier board optimised for the SST (including temperature sensors, optimal connectors and high-thermal conductivity). It should be noted that upgrading the SiPMs also results in changes to the FEE. 
  \item {\bf BP:} A design iteration of the BP is taking place to include the functionality of an additional board used in CHEC-S for data readout. 
  \item {\bf Calibration Flasher:} The calibration flasher has been iterated, replacing the previous design that included ten LEDs with a single LED of adjustable brightness. 
  \item {\bf Overall Mechanical Design:} The increase in enclosure size required to accommodate a flat entrance window results in a revised enclosure design. In addition the mechanical concept of all sub-systems has been revised on optimise the overall assembly and maintenance concept. 
\end{itemize}

\section{Integrated Telescope-Camera Verification}

Preliminary verification of the telescope structure, optics and camera concept was performed by installing the CHEC-S prototype camera on the ASTRI-Horn prototype telescope structure in April - July 2019. Over two observing campaigns the expected on-sky performance was verified as well as all interfaces, and basic functional and logistic issues (such as camera transport, turn-on tests, telescope installation procedure). 

\begin{figure}[t!]
    \centering
    \includegraphics[width=0.9\textwidth]{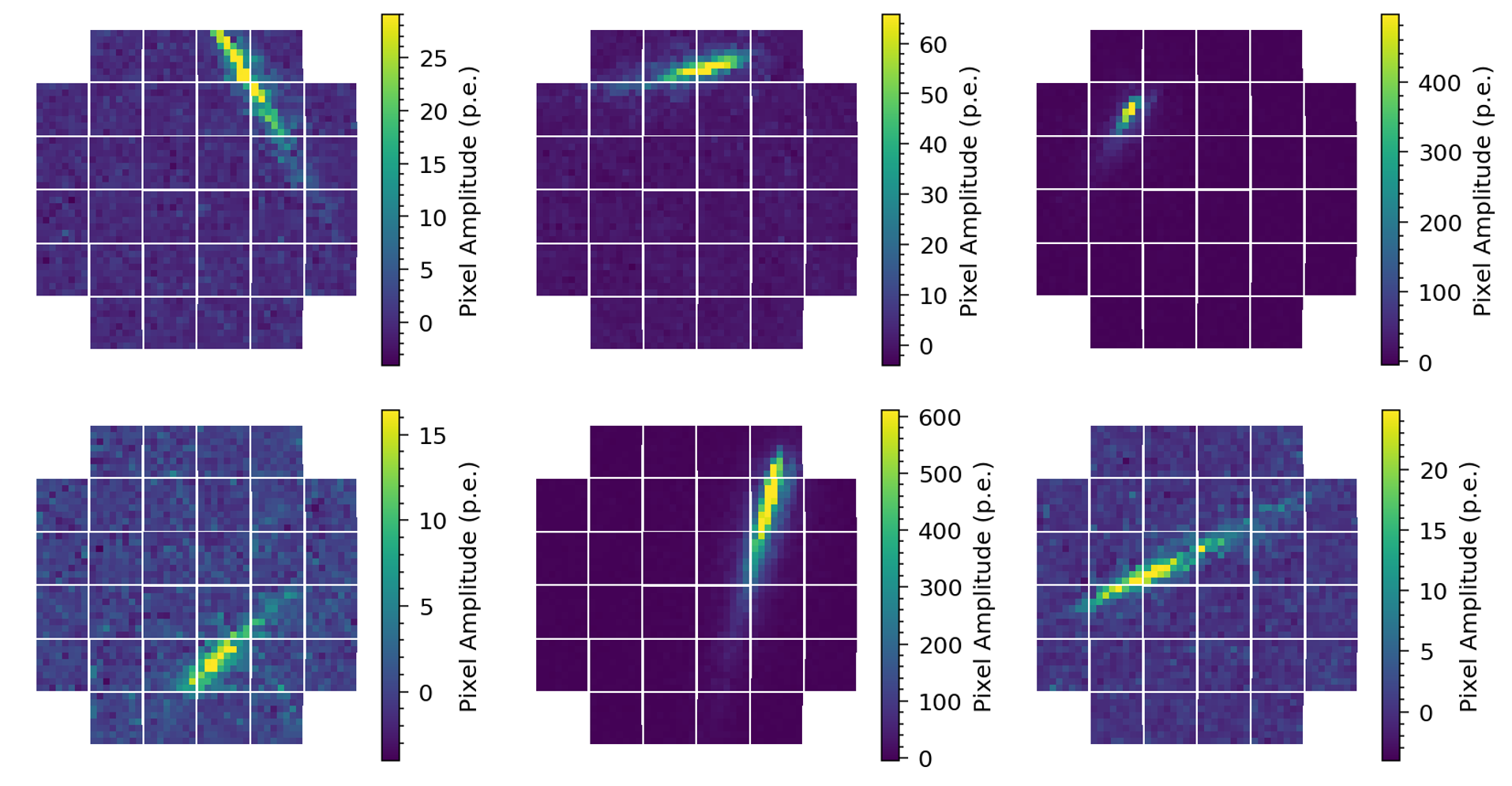}
    \caption{Examples of Cherenkov images recorded with CHEC-S on the ASTRI-Horn telescope.}
    \label{cherenkov}
\end{figure}

Following arrival on site, only a few hours were needed to verify camera functionality prior to installation on the telescope. Installation of the camera on the telescope required two people, a cherry picker, and $\sim$1 hour. The camera chiller, pipes and all cabling had been installed by the ASTRI team prior to the camera arrival following documentation provided by the camera team. Around one day of tests to verify the cooling system and camera functionality were performed before the first Cherenkov images were recorded soon after. The focus of both campaigns was engineering, to test and understand features of the camera in real conditions. Additionally, a limited number of hours (totalling $\sim$11.5 hours) were obtained for observing gamma-ray sources, including Mrk421 and 501. Comparison of extracted image parameters to Monte-Carlo simulations showed a good match, as did the behaviour of the trigger rates vs. threshold for varying NSB conditions. Figure~\ref{cherenkov} shows a selection of images captured with the camera during the campaigns. During observations recovery of the telescope pointing was also possible by using an additional feature of the camera. A second digitisation chain within the camera allows the SiPM signal integrated over 10 - 100 ms timescales to be read out. Astrometric calibration may then be performed by comparison to the expected star field. A similar procedure was used by the ASTRI team to verify the telescope structure and optics with the ASTRI camera and obtained comparable results~\cite{astri-pointing}.

\section{Design Finalisation and Series Production}

The SST Programme has been established within CTA to provide SSTs as an in-kind contribution to the Observatory. The programme will organise and oversee: the finalisation of the SST design; compliance with CTA interfaces and requirements; and plan for, then track, the series production phase. Design finalisation will conclude with a formal review to verify the design against CTA requirements and signal acceptance to proceed with production. 

During the series production of SSTs, structures and optics will by produced by commercial partners in Europe. The telescope structure and optics will undergo quality control before shipment to CTA-South under the strict supervision by INAF and OBP/CNRS. Cameras components will be procured, assembled into camera subsystems and tested by participating institutes. Full cameras will then be assembled from these subsystems and undergo functional tests at two production sites, with all cameras will passing through a single site for final pre-shipment verification tests. 

On-site at CTA-South, a well-trained team will commission telescope structures, which can be done independently to the cameras (using a temporary, purpose-built optical camera). Following basic in-coming control tests cameras will be stored on-site until a given telescope is ready for camera installation. As the camera is small and light, installation on telescope is fast. Extensive lab testing and preparation will be done off-site to allow efficient camera commissioning on telescope. A final set of formal acceptance tests for each fully integrated SST will be done prior to acceptance of the unit by the CTA Observatory. 

\section{Summary and Outlook}

The choice of underlying concept for the CTA SSTs has been made. The SST design will be based on the ASTRI and CHEC dual-reflector prototype structure and cameras, taking into account the previous experience gained on all designs. The ASTRI and CHEC prototypes have been shown to meet CTA requirements, individually and as an end-to-end SST system. Further development on both the camera and structure is underway. An SST Programme has been established to finalise the design and provide all required SSTs as an in kind contribution to the CTA Observatory.

\section{Acknowledgements}
This work was conducted in the context of the CTA SCT Collaboration.
We gratefully acknowledge financial support from the agencies and organizations listed here:
\url{http://www.cta-observatory.org/consortium}

\bibliographystyle{ICRC}
\bibliography{references}

\clearpage
\section*{Full Authors List: \Coll\ Consortium}

\scriptsize
\noindent
J.P.~Amans$^{1}$, 
D.~Berge$^{2}$, 
G.~Bonanno$^{3}$, 
R.B.~Bose$^{4}$, 
A.M~Brown$^{5}$, 
J.H.~Buckley$^{4}$, 
P.M.~Chadwick$^{5}$, 
F.~Conte$^{6}$, 
G.~Cotter$^{7}$, 
F.~de Frondat$^{8}$, 
N.~De Simone$^{2}$, 
J.L.~Dournaux$^{8}$, 
C.A.~Duffy$^{9}$, 
S.~Einecke$^{10}$, 
G.~Fasola$^{8}$, 
S.~Funk$^{11}$, 
G.~Giavitto$^{2}$, 
J.A.~Hinton$^{6}$, 
J.M.~Huet$^{8}$, 
N.~La Palombara$^{12}$, 
J.S.~Lapington$^{9}$, 
P.~Laporte$^{8}$, 
S.A.~Leach$^{9}$, 
G.~Leto$^{13}$, 
S.~Lloyd$^{5}$, 
S.~Lombardi$^{14}$, 
A.~Nayak$^{6}$, 
A.~Okumura$^{15}$, 
G.~Pareschi$^{16}$, 
H.~Prokoph$^{2}$, 
E.~Rébert$^{8}$, 
D.~Ross$^{9}$, 
G.~Rowell$^{10}$, 
S.~Scuderi$^{12}$, 
H.~Sol$^{8}$, 
S.~Spencer$^{7}$, 
H.~Tajima$^{15}$, 
A.~Trois$^{16}$, 
S.~Vercellone$^{16}$, 
J.~Vink$^{17}$, 
J.J.~Watson$^{2}$, 
R.~White$^{6}$, 
A.~Zech$^{8}$
\\
\\

\noindent
$^{1}$Department of Physics, Columbia University, 538 West 120th Street, New York, NY 10027, USA
$^{2}$Deutsches Elektronen-Synchrotron, Platanenallee 6, 15738 Zeuthen, Germany
$^{3}$INAF - Osservatorio Astrofisico di Catania, Via S. Sofia, 78, 95123 Catania, Italy
$^{4}$Department of Physics, Washington University, St. Louis, MO 63130, USA
$^{5}$Centre for Advanced Instrumentation, Dept. of Physics, Durham University, South Road, Durham DH1 3LE, United Kingdom
$^{6}$Max-Planck-Institut für Kernphysik, Saupfercheckweg 1, 69117 Heidelberg, Germany
$^{7}$University of Oxford, Department of Physics, Denys Wilkinson Building, Keble Road, Oxford OX1 3RH, United Kingdom
$^{8}$LUTH, GEPI and LERMA, Observatoire de Paris, CNRS, PSL University, 5 place Jules Janssen, 92190, Meudon, France
$^{9}$Dept. of Physics and Astronomy, University of Leicester, Leicester, LE1 7RH, United Kingdom
$^{10}$School of Physical Sciences, University of Adelaide, Adelaide SA 5005, Australia
$^{11}$Friedrich-Alexander-Universit\"{a}t Erlangen-N\"{u}rnberg, Erlangen Centre for Astroparticle Physics (ECAP), Erwin-Rommel-Str. 1, 91058 Erlangen, Germany
$^{12}$INAF - Istituto di Astrofisica Spaziale e Fisica Cosmica di Milano, Via A. Corti 12, 20133 Milano, Italy
$^{13}$INAF - Osservatorio Astrofisico di Catania, Via S. Sofia, 78, 95123 Catania, Italy
$^{14}$INAF - Osservatorio Astronomico di Roma, Via di Frascati 33, 00040, Monteporzio Catone, Italy
$^{15}$Institute for Space-Earth Environmental Research, Nagoya University, Chikusa-ku, Nagoya 464-8601, Japan
$^{16}$INAF - Osservatorio Astronomico di Brera, Via Brera 28, 20121 Milano, Italy
$^{1}7$Anton Pannekoek Institute/GRAPPA, University of Amsterdam, Science Park 904 1098 XH Amsterdam, The Netherlands

\end{document}